\documentclass[prl,showpacs,twocolumn]{revtex4}

\usepackage{amsmath}    

\usepackage{verbatim}   

\usepackage{color}      

\usepackage{hyperref}   

\usepackage{graphics}

\usepackage{epsfig}

\usepackage[english]{babel}

\usepackage{natbib}

\usepackage{tikz}
\usepackage{pdfpages}

\begin{document}

\title{Moderate stem cell telomere shortening rate postpones cancer onset in stochastic model.}

\author{Simon Holbek, Kristian Moss Bendtsen, Jeppe Juul}

\affiliation{University of Copenhagen, Niels Bohr Institute, Blegdamsvej 17, DK-2100 Copenhagen, Denmark}


\date{\today}

\begin{abstract}

Mammalian cells are restricted from proliferating indefinitely. Telomeres at the end of each chromosome are shortened at cell division and, when they reach a critical length, the cell will enter permanent cell cycle arrest - a state known as senescence. This mechanism is thought to be tumor suppressing, as it helps prevent precancerous cells from dividing uncontrollably.

Stem cells express the enzyme telomerase, which elongates the telomeres, thereby postponing senescence. However, unlike germ cells and most types of cancer cells, stem cells only express telomerase at levels insufficient to fully maintain the length of their telomeres leading to a slow decline in proliferation potential. It is not yet fully understood how this decline influences the risk of cancer and the longevity of the organism.

We here develop a stochastic model to explore the role of telomere dynamics in relation to both senescence and cancer. The model describes the accumulation of cancerous mutations in a multicellular organism and creates a coherent theoretical framework for interpreting the results of several recent experiments on telomerase regulation.

We demonstrate that the longest average cancer free life span before cancer onset is obtained when stem cells start with relatively long telomeres that are shortened at a steady rate at cell division. Furthermore, the risk of cancer early in life can be reduced by having a short initial telomere length.

Finally, our model suggests that evolution will favour a shorter than optimal average cancer free life span in order to postpone cancer onset until late in life.

\end{abstract}


\maketitle

\section{Introduction}

Telomeres are the non-coding ends of chromosomes that prevent loss of genomic information during chromosome replication \cite{Blasco:2005ix}. Each cell division leads to a telomere shortening of 50-100 base pairs, partly due to what is known as the end-replication problem \cite{Lorenz:2001,Huffman:2000wy}. When the telomeres reach a critical length the cell goes senescent, a state of permanent replication arrest that prohibits any further proliferation \cite{Campisi:2007jd}. The number of divisions a cell can undergo before it reaches senescence is called the Hayflick limit \cite{Campisi:2005do}.

Short telomeres have been linked to increased mortality and age related diseases \cite{Cawthon:2003bx} and accumulation of senescent cells is seen as one of the major causes of ageing \cite{Campisi:2005do, Collado:2007vx, bendtsen2012fragile}. However, the proliferation limit associated with telomere attrition is thought to work as a fail-safe mechanism against cells that divide in an uncontrolled and rapid fashion, particularly cancer cells \cite{Collado:2010vi}.

Genes that are critical in maintaining cells in a non-cancerous state are called proto-oncogenes or tumor suppressor genes. Estimates of the number of mutations in these genes needed for a healthy cell to turn cancerous range from 3-6 \cite{Alberts:2008vj} and above \cite{Beckman:2005wg}.

Cancer cells replicate uncontrollably and have unlimited proliferation capability, known as cellular immortality \cite{Hiyama:2007wy}. Consequently, the cancerous cell must by-pass the end-replication problem, which in 90\% of cancers is done by expressing the enzyme telomerase \cite{Shay:2012bw}. Telomerase, which mainly consists of the two components TERT and TERC, elongates the telomeres such that the telomere length of cancer cells is maintained during replication \cite{Blasco:2007wk}. Besides being present in cancer cells, telomerase is also found in stem cells and germ line cells \cite{Hiyama:2007wy}.


The amount of telomerase in germ line cells is sufficient to maintain telomere length \cite{Blasco:2005ix}. For stem cells, however, the level of telomerase is so low that telomeres are still shortened after each cell division, but at a lower rate than for somatic cells \cite{Cong:2002ts}.

It is an open question why stem cells express telomerase, and why the level of telomerase is not high enough to avoid a shortening of the telomere length throughout life.

It has been suggested that telomere shortening is a trade-off between limiting oncogenesis and reducing physiological ageing \cite{Campisi2001,Campisi:2005vh,Rodier:2011}.

The capability of cells to tune their telomere shortening rate by varying the expression of telomerase suggests the possibility of optimal telomere shortening strategies.

In this paper we present a stochasticl model that analyses how shortening of stem cell telomeres are coupled to the trade-off between ageing and postponing cancer. Previous mathematical models for telomere dynamics have mostly focused on the cellular transition to a senescent state \cite{Portugal:2008kh,opdenBuijs:2004dh}, and previous cancer models have focused mostly on oncogenesis \cite{Tomlinson:2002,Beckman:2006}. This work tries to bridge these areas of focus, developing a more coherent mathematical modelling framework for considering telomere dynamics in relation to both senescence and cancer. We find that there exists a telomere shortening rate that optimizes the cancer free life span of the organism. Our results also show that a suboptimal cancer free life span strategy may be selected in order to reduce the risk of cancer before the childbearing age.

\section{The model}

The stochatic model describes the accumulation of cancerous mutations in a multicellular organism, where each cell division will result in a shortening of the telomere sequence due to the end-replication-problem.

Mutations can arise during division of either somatic cells or stem cells. Initially an unmutated stem cell divides, thereby creating a somatic cell and a daughter stem cell. The somatic cell then proliferates until it, after $H_0$ divisions, reaches the Hayflick limit and undergoes senescence.

The Hayflick limit can be interpreted as the conversion of an initial telomere length, since a correlation between the replicative capacity of a cell and its initial telomere length has been shown \cite{Allsopp:1992}.

During each division, the somatic cell has a probability $p$ of acquiring a cancerous mutation, which is then propagated through the cell lineage in our model. In order to maintain homeostasis, only one daughter cell survives after each replication.

When the somatic cell has gone senescent it is removed and all mutations accumulated in the somatic cell lineage are lost.

The daughter stem cell then divides to produce a new daughter stem cell and a new generation of somatic cells.

Each time a stem cell divides, the Hayflick limit $H_G$ is reduced by a constant amount $\alpha > 0$, corresponding to the daughter stem cell having shorter telomeres than its parent. Additionally, there is a probability $p_{sc}$ of a cancerous mutation occurring in the self-renewing stem cell. Such mutations are permanent and are inherited by all stem cells in the stem cell lineage in the model.

If at any point the cell accumulates more than $C_m$ mutations, the organism is no longer cancer free. This will inevitably happen as mutations are steadily accumulated in the stem cell lineage, such that somatic cell lines start with more and more mutations.

\begin{figure}[tb]

\centering

\includegraphics[clip=true,,width=\columnwidth]{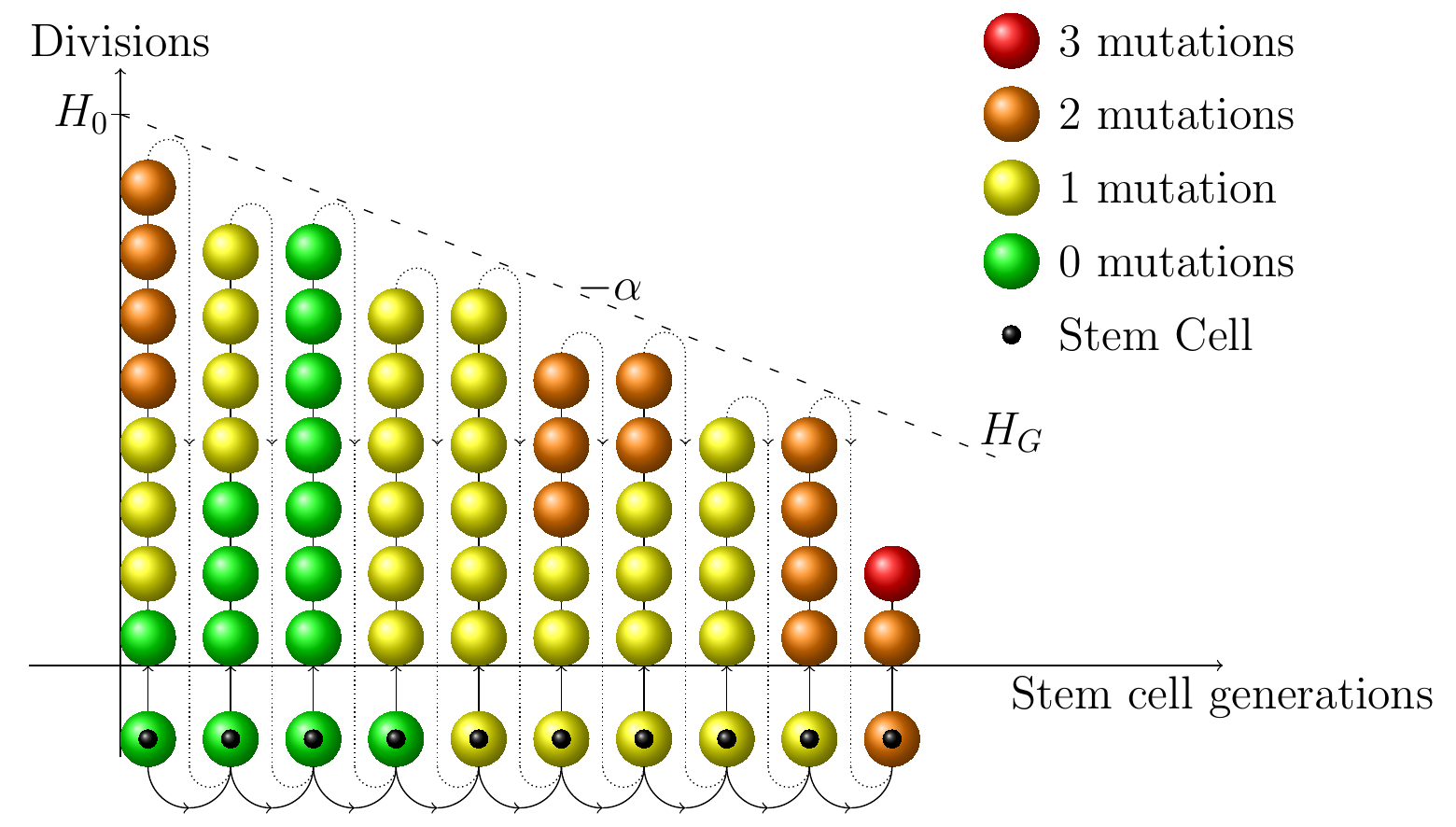}

\caption{Schematic illustration of the model. The flow of time follows the dashed arrows. Initially, a stem cell divides. One daughter cell specializes into a somatic cell, which proliferates until it reaches the Hayflick limit and goes senescent (first vertical row). The daughter stem cell then divides to produce a new daughter stem cell and a new generation of somatic cells (subsequent vertical rows). For each generation the Hayflick limit $H_G$ is reduced by a constant amount $\alpha$. At every cell division, somatic cells and stem cells have the probabilities $p$ and $p_{sc}$, respectively, to acquire cancerous mutations. Mutations in stem cells are permanent, while mutations in somatic cell lineages are lost when the cell line reaches the Hayflick limit. When the system has accumulated more than $C_m$ mutations (which in this illustration is 2), the organism will have developed cancer. The total cancer free life span is thus the sum of all somatic divisions.}

\label{fig1:model}

\end{figure}

The total number of cell divisions that the system undergoes before the onset of cancer is interpreted as the cancer free life span of the organism. A schematic illustration of the model is shown in Fig. \ref{fig1:model}.

\section{Theoretical analysis}

In this section we derive a mathematical expression for the average cancer free life span obtained for different telomere shortening strategies.

Each time a stem cell is introduced to the tissue, the Hayflick limit is reduced from its initial value $H_0$ due to the telomere shortening of the stem cell. After $G$ generations the Hayflick limit $H_G$ is given by

\begin{equation}
H_G = \lfloor H_0 - \alpha  G  \rceil   \label{eq:shortening}
\end{equation}

where $\lfloor \rceil$ denotes rounding to the nearest integer. During each of the subsequent $H_G$ cell divisions, the cells have a probability $p$ of acquiring a cancerous mutation. The probability $F(x)$ that the cell line acquires $x$ or fewer mutations before it reaches the Hayflick limit is given by the Binomial cumulative distribution function

\begin{equation}
F(x) = \sum_{i=0}^{x} {H_G \choose i} p^i (1-p)^{H_G-i} . \label{eq:cdf}
\end{equation}

At each stem cell division there is a risk $p_{sc}$ of acquiring a permanent mutation in the stem cell lineage. The probability for the organism to have $j$ mutations in the stem cells after $G$ generations, but still be cancer free, is denoted $S_{G,j}$. This is given by the recurrence relation

\begin{equation}
S_{G+1,j} =\big(  S_{G,j} (1-p_{sc}) + S_{G,j-1} p_{sc}  \big)  F(C_m-j) \label{eq:recurrence},
\end{equation}

where the expression in the bracket is the probability of having $j$ mutations in the stem cell, and $F(C_m-j)$ is the probability of not exceeding the critical number of $C_m$ mutations during the somatic divisions, thereby developing cancer.  

The initial stem cell is unmutated, $S_{0,0} = 1$, so in terms of Kroeniker-delta the starting condition is

\begin{equation}
S_{0,j} = \delta_{0j}  \label{eq:S0}
\end{equation}

Using Eq. \eqref{eq:recurrence} and Eq. \eqref{eq:S0}, the system can be solved numerically. Notice that when $\alpha >0$, the Hayflick limit will eventually drop to zero, after which the stem cell undergoes senescence even if the cell line is still cancer free. According to Eq. \eqref{eq:shortening} this will happen after $\frac{H_0 - 0.5}{\alpha}$ generations, which therefore gives an upper bound for the total cancer free life span of the organism.

The overall probability that the organism will not have developed cancer at a given generation is defined as.

\begin{equation}
S_G=
\begin{cases}
\sum_j S_{G,j} & \text{for } G\le \frac{H_0 - 0.5}{\alpha} \\
0 & \text{for } G>\frac{H_0 - 0.5}{\alpha}
\end{cases}
\label{eq:SurvivalProb}
\end{equation}

The average cancer free life span $\langle L (H_0,\alpha) \rangle$ is the mean number of cell divisions the organism will live before exceeding the critical number of mutations. It can be found by multiplying the probability of getting cancer at each generation $(S_G-S_{G+1})$ by the number of cancer free generations and the average Hayflick limit each of these consisted of.

\begin{equation}
\langle L (H_0,\alpha ) \rangle = \sum_{G=1}^{\infty} (S_G-S_{G+1}) G  \frac{H_0 + H_G}{2} \label{eq:LifeSpanVarying}
\end{equation}

When $\alpha = 0$ the Hayflick limit $H_0$ is constant and Eq. \eqref{eq:LifeSpanVarying} becomes

\begin{equation}
\langle L (H_0, 0) \rangle=H_0\sum_{G=1}^{\infty} S_G \label{eq:LifeSpanConst}
\end{equation}

Note that for this case $S_G$ will never reach 0 and a cut off is necessary. This is made when $S_G<10^{-5}$  and has negligible influence on the cancer free life span.

\section{Results}                  

We now fix the ratio between the mutation probabilities such that $\frac{p}{p_{sc}} = 100$ according to \cite{Albertini:1990wg,Cervantes:2002ul}, and set $p$ = 10$^{-2}$ and $C_m = 6$.

These choices of parameter values are not essential for the conclusions and other choices yield qualitatively similar results (see supplementary material).

\begin{figure}[tb]

\centering

\includegraphics[width=\columnwidth]{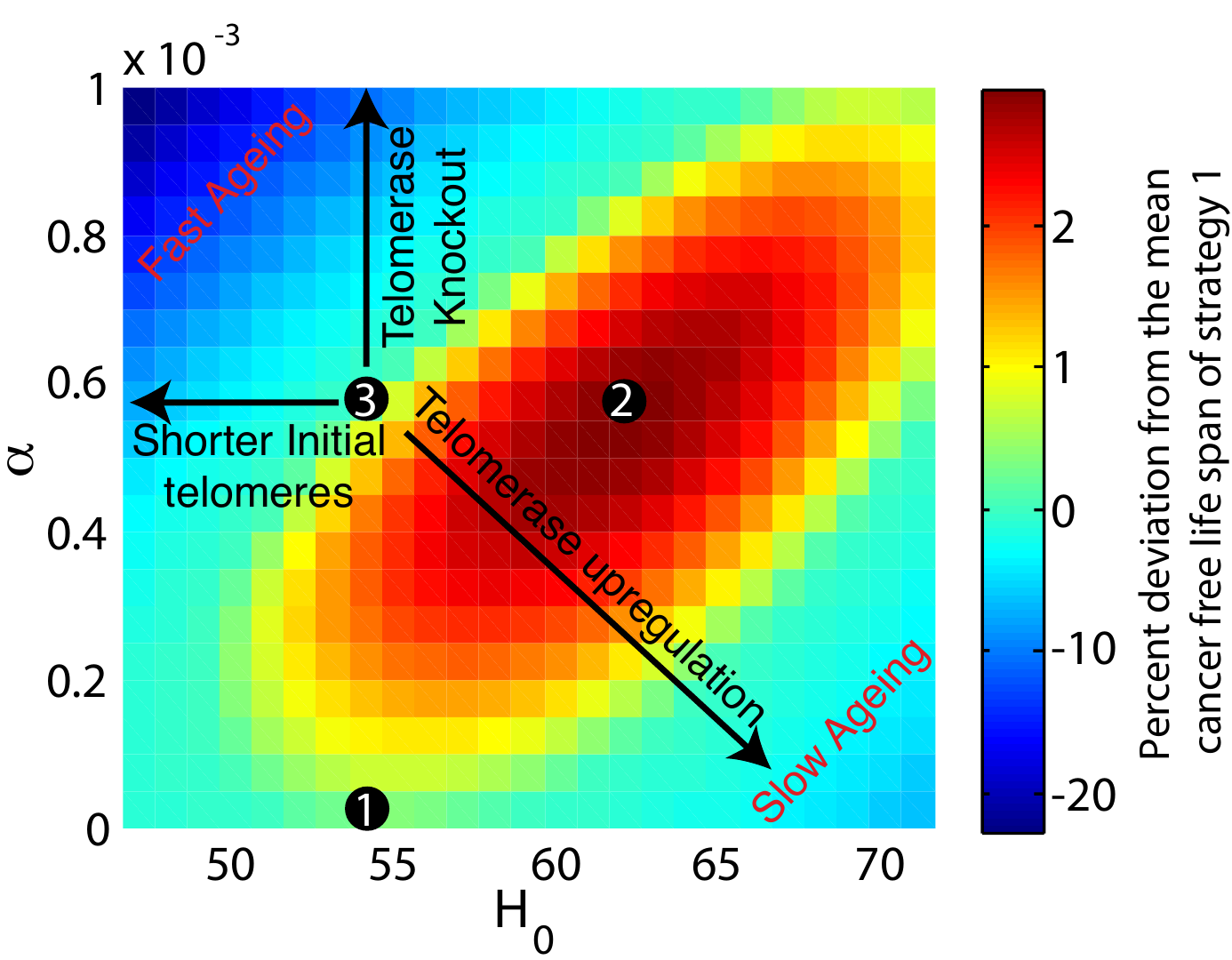}
\caption{Deviation in average cancer free life span for different initial Hayflick limits and shortening factors $\alpha$ compared to the longest possible cancer free life span for constant Hayflick limit (strategy 1). The overall longest cancer free life span (strategy 2) is obtained for a higher initial Hayflick limit, which is then slowly reduced through telomere shortening in stem cells. Strategy 3 has the same initial Hayflick limit as strategy 1 and the same telomere shortening rate as strategy 3. Black arrows show how different experimental setups can change the position in parameter space through modifications of the initial Hayflick limit of stem cells and the telomere shortening rate. A low initial Hayflick limit and high telomere shortening rate cause premature ageing due to increased accumulation of senescent cells.
Parameters: $C_m$ = 6, $p=10^{-2}$, $p_{sc}=10^{-4}$.  }
\label{fig:AlfaBeta}

\end{figure}

The average cancer free life span is calculated from Eq. \eqref{eq:LifeSpanVarying} and Eq. \eqref{eq:LifeSpanConst} and their performance compared to the strategy with no stem cell telomere shortening that yields the longest cancer free life span, as obtained from Eq. \eqref{eq:LifeSpanConst}. The result is shown in Fig. \ref{fig:AlfaBeta}, where three strategies for further study are labeled in the following way:

\begin{enumerate}

\item The strategy where stem cells express a telomerase activity, such that the telomere length of stem cells is constant over time ($\alpha=0$). The initial Hayflick limit is chosen such that the cancer free life span is maximized.

\item The strategy that gives the longest cancer free life span when allowing stem cell telomeres to shorten at each division ($\alpha>0$). Again, the initial Hayflick limit is chosen such that the cancer free life span is maximized. This results in a larger initial Hayflick limit than for strategy 1.

\item The strategy that combines the initial Hayflick limit from strategy 1 and the stem cell telomere shortening rate from strategy 2. Note that this strategy yields similar cancer free life span as obtained for strategy 1.

\end{enumerate}



The shorter the telomere length of stem cells is, the more frequent is the need for self renewal in order to maintain homeostasis in the tissue. Every self renewal is accompanied with the risk of accumulating an additional permanent mutation \cite{Cervantes:2002ul}. Therefore, a lower value for $H_0$ increases the rate at which mutations occur in the stem cells.

Furthermore, a shorter $H_0$ will reduce the amount of cell divisions before stem cell senescence, resulting in faster ageing and a shorter maximum cancer free life span.

Keeping $H_0$ fixed and down-regulating the telomerase production is equivalent to increasing $\alpha$, thus moving upwards in Fig. \ref{fig:AlfaBeta}. The higher $\alpha$ will suppress the risk of developing cancer, as the risk can be managed by undergoing fewer divisions before senescence. The higher turnover rate results in systems where cancer is rarely developed, but where the cancer free life span of otherwise healthy cell lineages is limited by stem cells going senescent.

Up-regulation of telomerase in stem cells results in a lower $\alpha$ while up-regulation of telomerase in somatic cells will increase the proliferation potential, which is equivalent to a higher $H_0$. These two effects combined  can, therefore, in Fig. \ref{fig:AlfaBeta} be illustrated as a shift downwards to the right.

This shift of strategy can in small quantities be advantageous and prolong life, but is also associated with an increased risk of developing cancer since somatic cells are able to continue proliferating for longer before the senescent fail-safe mechanism sets in.

\begin{figure}[tb]

\centering

\includegraphics[width=\columnwidth]{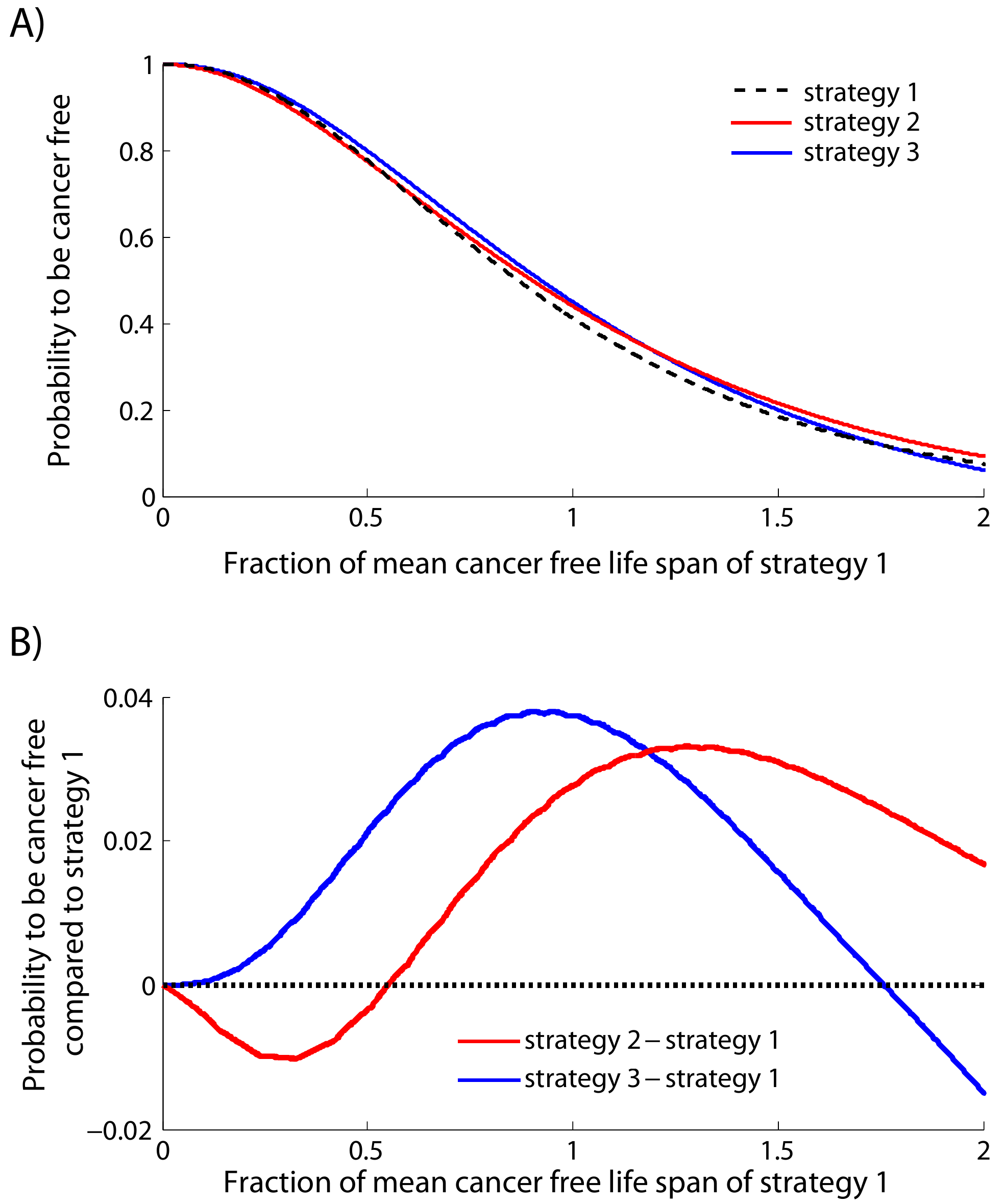}

\caption{Probability of not yet having developed cancer as a function of time for different telomere shortening strategies. \textbf{A:} For all three strategies, the probability slowly decreases with time. Only very few individuals will develop cancer later than twice the mean cancer free life span of the population.
\textbf{B:} Probability of not yet having developed cancer for strategy 2 and 3 compared to strategy 1. Strategy 2, which has the longest average cancer free life span, will have a higher probability of cancer development early in life. Strategy 3, which has a low initial Hayflick limit, is able to postpone cancer until later in life.}
\label{fig:Versus2}

\end{figure}

These observation are supported by Fig. \ref{fig:Versus2}a and \ref{fig:Versus2}b, which show the probability of the organism to still be cancer free as a function of time, for each of the strategies 1, 2, and 3.

Surprisingly, we see that strategy 2, which yields the longest cancer free life span, is also the one that is most exposed to cancer early in life. This strategy benefits from less stem cells attrition late in life, thereby supplying the system with less mutated cells, which in the end increases the average time before cancer development. Minimizing the risk of developing cancer at an early age can be done by lowering $H_0$, raising $\alpha$, or both. With strategy 3 a higher percentage of a population would therefore still be a cancer free in the early in life compared to other strategies as 1 and 2 that have similar or longer cancer free life span.

These results indicate that evolution would favour species with a restricted telomerase production in their stem cells above species that maintain their stem cells telomere length fully throughout life. This shortening rate can then be adjusted such that a lower cancer risk is obtained at the childbearing age, thereby increasing the chance of giving birth.

Therefore, strategies that solely optimize cancer free life span may not be advantageous from an evolutionary point of view, if they increase the risk of cancer early in life when reproduction takes place, as seen with strategy 2. We also find that cancer free life span for some strategies, e.g. Strategy 3, can be prolonged even further if telomerase is up-regulated at a later stage in life (see supplementary materials).

Strategy 3 and other stem cell shortening strategies are not without downsides, as the somatic cells will undergo senescence after fewer and fewer divisions. Shorter telomeres and an accumulation of senescent cells are closely related to biological ageing \cite{Cawthon:2003,Herbig:2006}, hence, strategies that suppresses cancer early in life by shortening their telomeres are associated with accelerated aging.

\section{Discussion and conclusion}

In our model a decrease of telomerase activity (TERT or TERC) leads to a faster shortening of stem cell telomeres per cell division. Our model predicts that any reduction of telomerase activity decreases the risk of cancers early in life, but at the same time causes a higher rate of stem cell replications. The increased replication rate leads to an increased rate of cancerous mutations in the stem cells. For certain regimes the advantages of telomerase activity down regulation resulting in a low risk of cancer early in life is outweighed by the higher risk of cancer later in life, and therefore the total cancer free life span of the organism is reduced. The increased cancer risk when telomerase is removed was observed in a study by González et al. (2000) , where they found a slight increase in oncogenesis in highly proliferating tissue for TERC deficient mice\cite{GonzalezSuarez:2000hl}. In addition our model also predicts that stem cells will go senescent faster, and therefore cause faster ageing. This is 
also consistent with experimental observations; in a TERC mouse study Liu et al. (1998) found that a lack of telomerase caused the mice to present premature ageing symptoms, especially in highly proliferating tissue \cite{Lee:1998hb}.


Our model demonstrates that telomerase up-regulation, in general, leads to an organism more prone to developing cancer, which was observed experimentally by González et al. (2002)\cite{GonzalezSuarez:2002vl}.

Furthermore, the model predicts that a temporal induction of telomerase could increase cancer free life span (see Supplementary). If the organism starts with a moderate telomerase activity and then increases the activity, the organism could prevent cancer onset at an early stage of life and postpone ageing, thereby increasing longevity. A recent study by Bernandes de Jesus et al. (2012) supports this result. By inducing telomerase activity in adult mice, they were able to increase longevity without increasing the risk of cancer \cite{BernardesdeJesus:2012kq}.

Additionally, it has been shown that, when comparing two types of TERC deficient mice with different genetic backgrounds - one type with initially shorter telomeres and one with normal initial telomere length - the mice with shorter telomeres show decreased viability, not due to cancer, but due to premature ageing \cite{Rudolph:1999,Herrera:1999ba}. Our model demonstrates that decreasing $H_0$ leads to a faster attrition of the stem cell pool and therefore faster ageing.

In conclusion, our model offers an explanation to why stem cells express telomerase at a moderate level, such that their telomere lengths are not conserved but is shortened at a lower rate than in somatic cells. Furthermore, our results suggest that, in stem cells, evolution will favor short initial telomeres that are slowly shortened at each division (strategy 3), as this strategy is able to reduce the likelihood of cancer early in life.

\newpage
\onecolumngrid
\noindent
\newpage
\begin{center}
\textbf{-- SUPPLEMENTARY MATERIAL --}\
\end{center}

\vspace{1cm}
In this supplementary note, we show that the choices of the parameter values $C_m,p$ and $p_{sc}$ are not essential for the overall conclusions drawn in the article
regarding cancer free life span obtained by shortening the Hayflick limit after each stem cell division.

\maketitle
\section{}
\begin{figure}[h!]
 \includegraphics{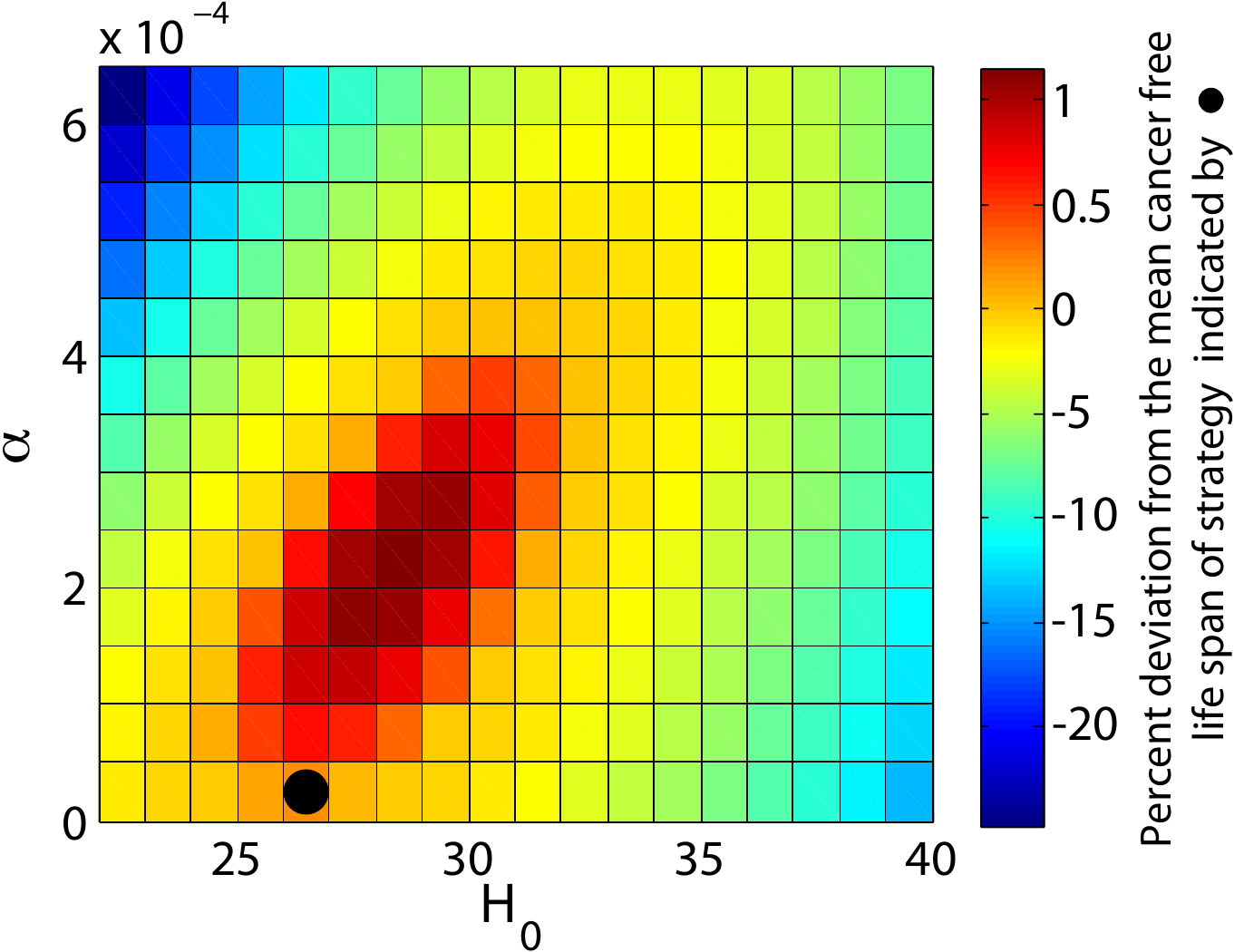}
\caption{Deviation in average cancer free life span for different initial
Hayflick limits and shortening factors $\alpha$ compared to the
longest possible cancer free life span for constant Hayflick limit (black dot).
Decreasing $C_m$ reduces the benefit one can obtain by shortening the stem cell telomeres in order to maximize
the cancer free life span compared to the strategy with constant Hayflick limit. Parameters: $C_m=4$, $p=10^{-2}$ and $p_{sc}=10^{-4}$.}
\label{fig:Experiment1}
\end{figure}

We here present three cases where the value of one parameter is varied, compared to those used for generating figure 2 in the article, while the two others kept constant.
We show again, that some shortening strategies can both prolong life and postpone cancer early in life, and that
the benefits of this depends on the choice of parameter values. The black dot on the figures represents the strategy with a constant Hayflick limit that produces
the longest cancer free life span.
First we keep $p$ and $p_{sc}$ fixed and change $C_m$ to 4.  See figure \ref{fig:Experiment1}

\begin{figure}[h!]
 \includegraphics{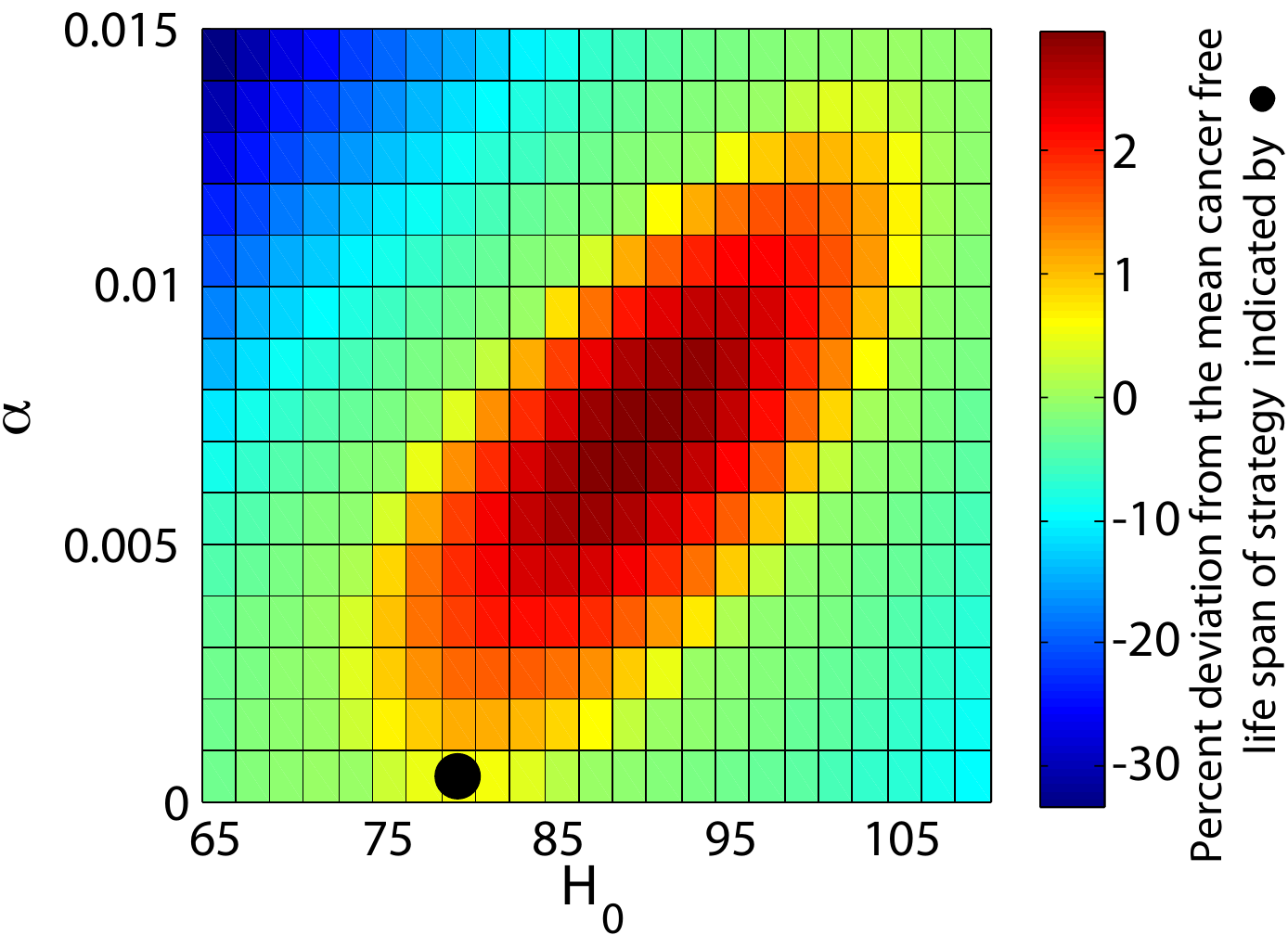}
\caption{Keeping $C_m$ and $p$ fixed, but increasing $p_{sc}$ by a factor 10, results in optimal strategies with higher $H_0$ and $\alpha$ as stem cells
should be introduced less frequent due to their increased mutation probability. Parameters: $C_m=6$, $p=10^{-2}$ and	 $p_{sc}=10^{-3}$.}
\label{fig:Experiment2}
\end{figure}

\newpage
Second, the stem cell mutation rate $p_{sc}$ is increased by a factor 10, with $C_m$ and $p$ kept fixed. See figure \ref{fig:Experiment2}

\begin{figure}[h!]
 \includegraphics{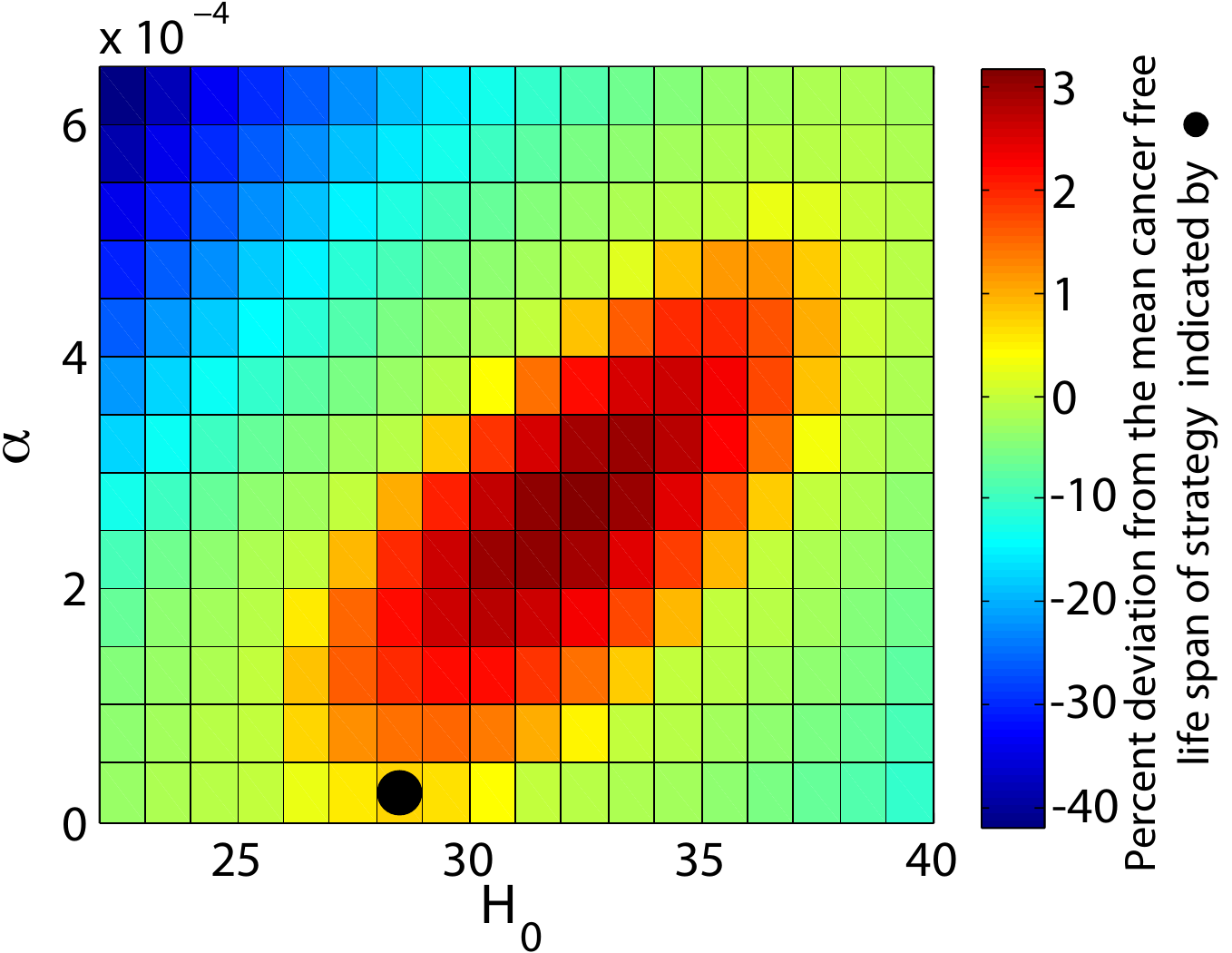}
\caption{Increasing the mutations rate for somatic cells will shift the optimal strategies towards lower $H_0$ and $\alpha$ as stem cells should now be introduced more frequently. Parameters: $C_m=6$, $p=2\cdot10^{-2}$ and $p_{sc}=10^{-4}$.}
\label{fig:Experiment3}
\end{figure}

\newpage
Lastly, the mutation rate for somatic cells $p$ is increased by a factor 2, with $C_m$ and $p_{sc}$ kept unchanged. See figure \ref{fig:Experiment3}

\begin{figure}[h!]
 \includegraphics[width=0.9\textwidth]{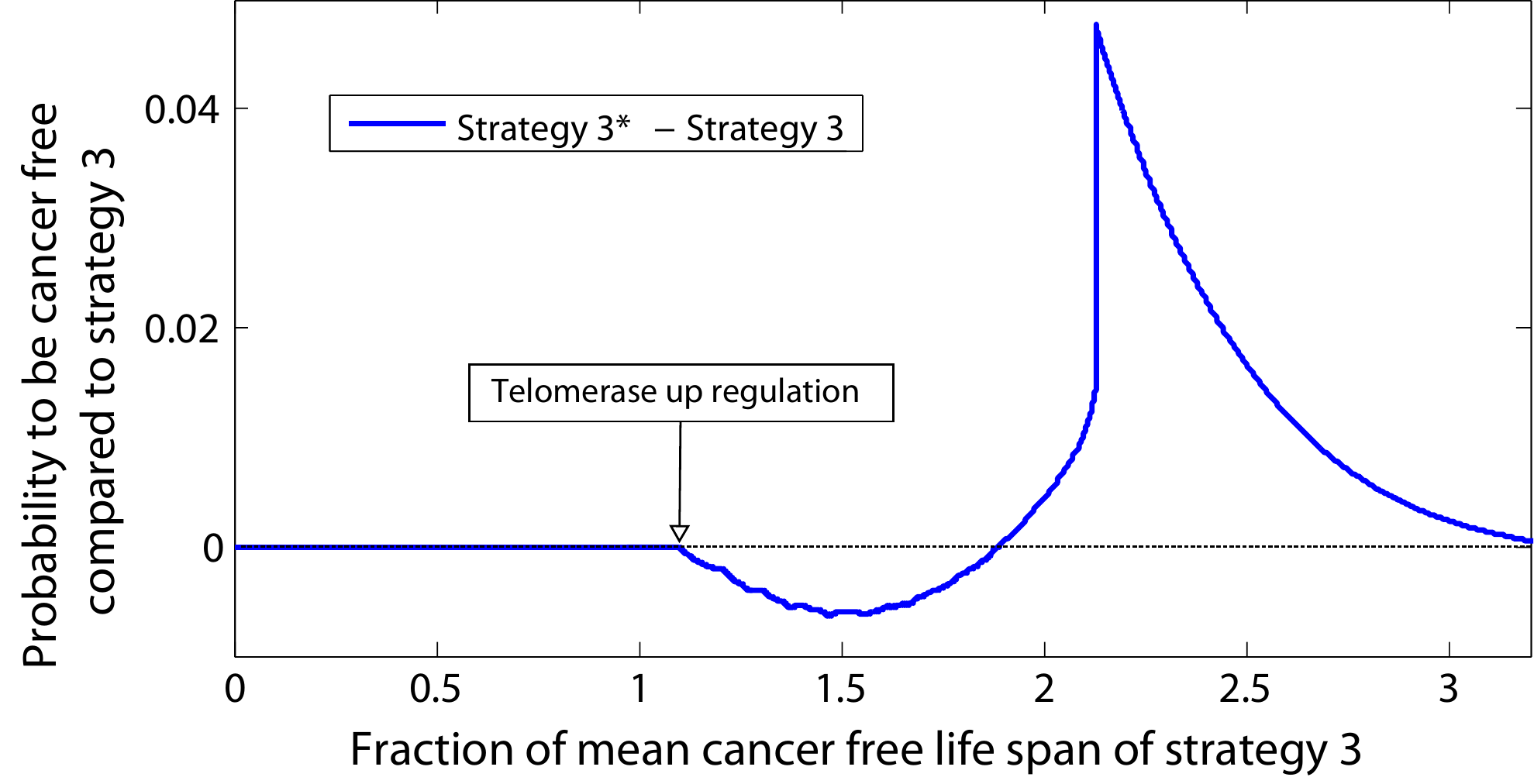}
\caption{Increased probability to be cancer free for two strategies with same initial conditions but with strategy 3* having an up-regulation of telomerase later in life. The up-regulation
(lower $\alpha$) leads to slightly increased cancer risk, but the strategy turns advantageous later on, as the increased telomerase production postpones the time at which
the stem cells go senescent. The stem cells of strategy 3 go senescent shortly after a time 2.1 longer than the mean cancer free life span, whereas this is postponed for strategy 3*.}
\label{fig:Figure4}
\end{figure}

\newpage
In order to verify that life can be prolonged by adding telomerase later in life, we here present one example of how this could occur. In this example we compare strategy
3 with strategy 3*. Strategy 3* has the same initial conditions as strategy 3, but will, when reaching $H_G$ = 39, double its telomerase production (halving $\alpha$)
for the remaining cancer free life span. With this up-regulation in telomerase production the longevity can be increased. The increased cancer free life span relies solely on the ability to postpone
stem cell senescence by stimulating it with telomerase. The steep peak in Fig. (\ref{fig:Figure4}), which is seen after a time 2.1 longer than the mean cancer free life span, is caused by the stem cells of
strategy 3 going senescent so $S_G$ immediately drops to 0.

\end{document}